# Revealing Local Structures through Machine-Learning-Fused Multimodal Spectroscopy

**Haili Jia[1,2], Yiming Chen[1,2], Gi-Hyeok Lee[3], Jacob Smith[4], Miaofang Chi[2,3,4], Wanli Yang[3], Maria K. Y. Chan[1,2*]**

[1] Center for Nanoscale Materials, Argonne National Laboratory, Woodridge, IL 60517, USA

[2] Energy Storage Research Alliance, Argonne National Laboratory, 9700 South Cass Avenue, Lemont, IL 60439, USA

[3] Advanced Light Source, Lawrence Berkeley National Laboratory, Berkeley, CA 94720, USA.

[4] Center for Nanophase Materials Sciences, Oak Ridge National Laboratory, Oak Ridge, TN 37831, USA

[5] Thomas Lord Department of Mechanical Engineering and Materials Science, Duke University, Durham, NC 27708, USA

* Corresponding author. Email address: mchan@anl.gov.

## Abstract

Atomistic structures of materials offer valuable insights into their functionality. Determining these structures remains a fundamental challenge in materials science, especially for systems with defects. While both experimental and computational methods exist, each has limitations in resolving nanoscale structures. Core-level spectroscopies, such as x-ray absorption (XAS) or electron energy-loss spectroscopies (EELS), have been used to determine the local bonding environment and structure of materials. Recently, machine learning (ML) methods have been applied to extract structural and bonding information from XAS/EELS, but most of these frameworks rely on a single data stream, which is often insufficient. In this work, we address this challenge by integrating multimodal *ab initio* simulations, experimental data acquisition, and ML techniques for structure characterization. Our goal is to determine local structures and properties using EELS and XAS data from multiple elements and edges. To showcase our approach, we use various lithium nickel manganese cobalt (NMC) oxide compounds which are used for lithium ion batteries, including those with oxygen vacancies and antisite defects, as the sample material system. We successfully inferred local element content, ranging from lithium to transition metals, with quantitative agreement with experimental data. Beyond improving prediction accuracy, we find that ML model based on multimodal spectroscopic data is able to determine whether local defects such as oxygen vacancy and antisites are present, a task which is impossible for single mode spectra or other experimental techniques. Furthermore, our framework is able to provide physical interpretability, bridging spectroscopy with the local atomic and electronic structures.

# Introduction

Understanding the local structure in materials is crucial for advancing various scientific and technological fields. From the electronic and thermal transport properties in thermoelectric materials to the functionality of ferroelectric compounds, local structure plays a pivotal role in unleashing the potential of materials for a wide range of applications.[1-5] This is particularly evident in nanoscale materials, where materials are engineered at the atomic or molecular scale, the understanding of local structure becomes even more crucial. Nanoscale interactions, defects, and the overall arrangement of atoms significantly impact the quality, reliability, and performance of these materials.[6-9] For instance, in energy storage and conversion technologies, subtle changes in local structures can transform materials' behavior, such as turning otherwise inert transition metal oxides into highly efficient electrocatalysts.[10] In multi-component battery cathodes, optimizing the composition and local structural features significantly improves battery performance, underscoring the need for precise control of local atomic arrangements. Subtle variations in atomic coordination can significantly affect the trade-off between capacity and long-term stability, where certain configurations may lead to improved electrochemical performance but also increase susceptibility to material degradation.[11,12]

Core-level spectroscopies, such as X-ray Absorption Spectroscopy (XAS), X-ray Photoelectron Spectroscopy (XPS), X-ray Emission Spectroscopy (XES), and Electron Energy Loss Spectroscopy (EELS), have become essential in revealing these local atomic and electronic environments. These techniques rely on core electrons' interactions with external probes, such as X-rays or electrons, to extract detailed information about chemical states, atomic coordination, and electronic structures. By measuring the energy lost or absorbed by core electrons during excitation, core-level spectroscopies provide highly sensitive, element-specific data that are invaluable for understanding subtle changes in materials, especially in complex and nanostructured systems where small variations can drastically alter performance.

Among these techniques, EELS and XAS stand out for their ability to probe local electronic structures, chemical states, and atomic coordination by analyzing core-electron excitations. EELS provides atomic-scale spatial resolution and allows for the detailed mapping of both chemical and electronic states at the nanoscale, making it ideal for studying local environments in nanomaterials. XAS, on the other hand, is particularly effective in in situ or operando conditions, enabling researchers to monitor real-time structural changes in materials during chemical reactions, phase transitions, or under operational conditions. This capacity to track changes as they occur in a material's working environment is crucial for optimizing materials in practical applications.[13-16]

However, characterizing local structures, especially in complex materials, presents significant challenges. Traditionally, the field has relied on a combination of experimental and computational methods to analyze these structures. While experimental techniques provide direct observations, they often struggle with overlapping spectral features, resolution limits, and sensitivity issues at the nanoscale where fine structural variations are critical.[8,17-19] On the other hand, computational approaches, including various forms of simulations, provide a complementary perspective. However, they too frequently encounter hurdles, such as the need for extensive computational resources and potential inaccuracies in modeling complex interactions, especially in defect-rich materials.[20-23]

In recent years, machine learning (ML) has emerged as a transformative tool in materials characterization, providing new pathways to overcome some of the limitations faced by traditional methods. ML algorithms have demonstrated remarkable capabilities in assisting the deconvolution of overlapping experimental spectral features, predicting material properties such as oxidation states, and identifying chemicals from absorption edges in an automated fashion.[17,24-26] Additionally, ML algorithms have also been applied to data synthesis and augmentation, helping to overcome limitations posed by data scarcity.[25,27-30] Despite these advancements, current frameworks in materials science often rely on single data sources, leading to potentially incomplete or biased understandings of complex materials. This is particularly evident in complex materials where multiple elements and interactions are involved, making it difficult for a single technique to capture all relevant information. For example, in multi-component cathodes like Nickel-Manganese-Cobalt oxides (NMC), the interactions between various elements and their local environments determine redox sensitivity, capacity contributions, and stability. However, this crucial information cannot be fully captured by focusing on a single element or using a single method.[11]

To address these challenges, this work proposes an effective and efficient approach that synergistically integrates *ab initio* simulations and advanced machine learning techniques on multimodal characterization. By overcoming the limitations of single-method analyses and leveraging the power of multimodal data, this work aims to provide a more comprehensive understanding of atomic structures, paving the way for the development of novel materials with tailored functionalities and enhanced performances.

We used NMC oxides as our sample system, as they have emerged as a cornerstone in the field of energy storage technologies, particularly in lithium-ion batteries. This prominence is due to their balanced properties, such as high energy density, relatively low cost, and environmental friendliness.[31] The quest for high-performance energy storage solutions has increasingly turned to NMC materials, which are crucial for a wide range of applications, from portable electronics to electric vehicles and grid storage.[32] However, the practical application and further development of NMC-based energy storage systems face challenges related to structural instability under operational conditions. These include phase transitions, cation mixing, and surface reconstruction, which significantly affect the long-term cycling stability and safety of batteries.[33] Given the significant impact of NMC materials on the future of energy storage, addressing their structural instability through comprehensive characterization and analysis is not only a scientific endeavor but also a technological imperative. Thus, we aim to use the insights gained from multimodal characterization to optimize NMC materials, thereby pushing the boundaries of energy storage technologies towards higher efficiency, capacity, and safety. This workflow is presented in **Figure 1**. Noted, the proposed framework is designed with versatility and general applicability, enabling its transfer and adaptation to various material systems, thus promising widespread utility across various fields of materials science and engineering.

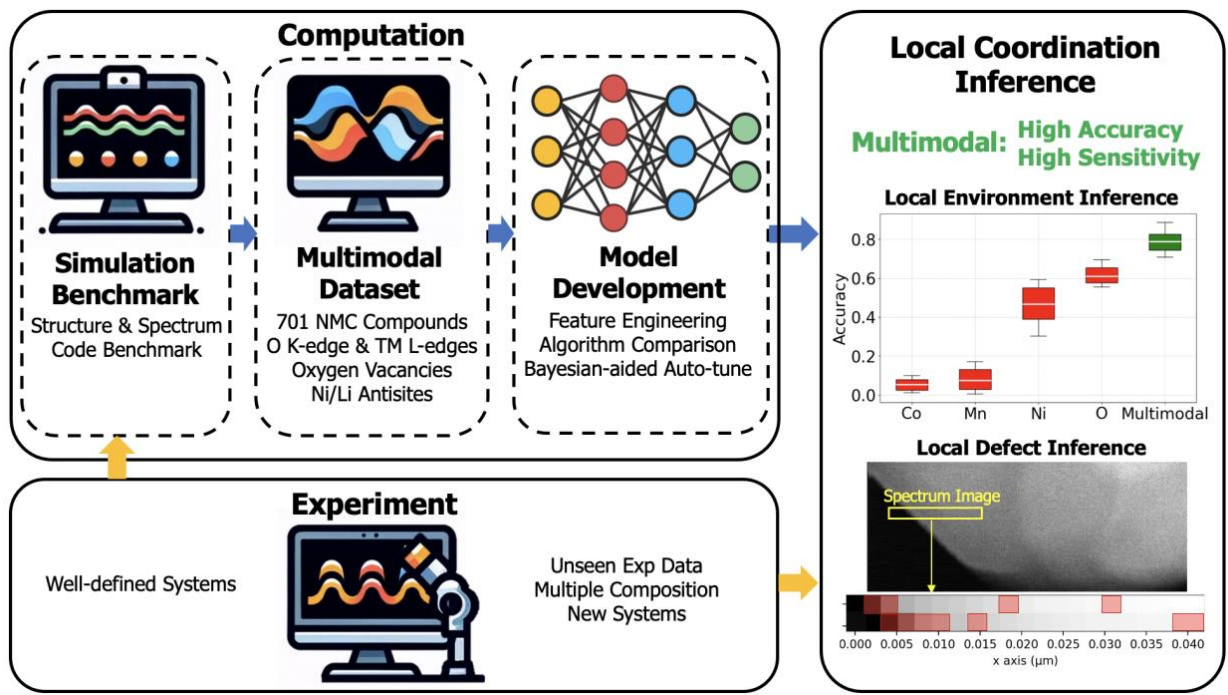

**Figure 1.** Overview of the workflow. The multimodal workflow is indicated by red arrows, and the local coordination inference results are shown in red. For comparison, the unimodal workflow is depicted by black arrows and gray results.

## Results and Discussion

### Structures and Spectra

We calculated 851 NMC structures, encompassing both pristine and defective configurations. Specifically, there are 701 pristine NMC structures, covering the entire lithiation spectrum from fully lithiated to fully delithiated states. Within this subset, 115 structures correspond to NMC811, 312 to NMC721, and 274 to NMC622. Additionally, 150 structures include defects - 136 with oxygen vacancies and 14 with Ni/Li antisites. Representative optimized structures are illustrated in **Figure 2(a-c)**. For each structure, we computed O K-edge and Ni/Mn/Co $L_{2,3}$-edges across all sites. We performed benchmarking tests with various codes and parameters, detailed in the **Methods** section. As a showcase, **Figure 2(d)** visualizes the raw simulated spectra for all pristine NMC811 configurations.

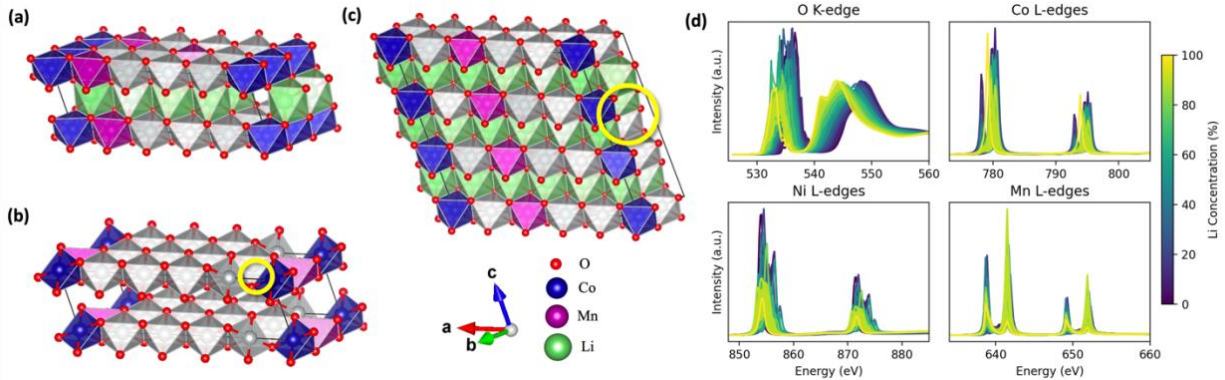

**Figure 2.** Representative optimized NMC structures: (a) Pristine NMC622 with 70% lithiation ($Li_7Ni_6Mn_2Co_2O_{20}$), (b) NMC721 with 5% oxygen defect ($Ni_7Mn_2Co_1O_{19}$), and (c) NMC811 with 3% Ni/Li antisites ($Li_{10}Ni_8Mn_1Co_1O_{20}$). Defect regions are highlighted with yellow circles. (d) Raw simulation data for NMC811 with 0-100% lithiation.

## **Machine Learning Model**

To apply machine learning models trained on simulation data to experimental data, manual adjustments to the simulated data are often required for better alignment. In our case, this calibration process includes shifting and applying appropriate broadening to the simulated data. We found that after calibration, using raw intensity data without dimensionality reduction yielded the best results. However, in cases where manual alignment was not performed, CDF outperformed all other features, consistent with findings by Chen et al. (2023). This robustness of CDF against shifts is particularly important for diverse experimental setups, where energy onsets may vary due to instrumental settings.[34,35] Thus, despite the slightly better performance of using raw intensity with manual alignment (approximately 2-5% higher accuracy, varying by task type), the resilience of CDF against shifts makes it a valuable asset. In pursuit of an automated framework applicable to experimental datasets, we therefore chose to use CDF as the input feature. Featurization and dimension reduction details are provided in the **Methods** section.

Among various ML models evaluated, tree-based methods delivered superior performance compared to other model types, aligning with findings from previous ML studies on core-loss spectra.[24,69] Within this category, XGBoost delivered the most stable and optimal performance. Gradient boosting machine (GBM) and LightGBM tended to overfit for local TM content inference due to their level-wise tree growth. Benchmarking results for these models are presented in the **Methods** section.

Training ML models is a time-intensive process that significantly impacts outcomes. To enhance efficiency and efficacy, we employed Bayesian Optimization (BO) for hyperparameter tuning. Unlike grid or random searches, BO uses historical results to inform decisions on the next set of hyperparameters to evaluate. This approach typically requires fewer iterations to identify optimal or near-optimal hyperparameters, which is particularly beneficial in scenarios with limited data or computational resources. By minimizing the number of evaluations, Bayesian Optimization optimizes the model-tuning process, make it particular useful for cases with limited data points. While Gaussian processes are commonly used as the surrogate model for Bayesian Optimization, scalability issues arise due to their cubic computational complexity with the number of data points.[36] To address this, we utilized the Tree-structured Parzen Estimator (TPE) for automated hyperparameter optimization, with algorithm details described in the **Methods** section.

In summary, we selected the CDF as the input feature and used the XGBoost model, optimized via BO-TPE, for all the tasks in this study. In the following subsections, we evaluated the characterization performance of the model and delved into the scientific interpretations underlying its functionality. By comparing the multimodal and unimodal approaches, we aimed to determine which approach offers more comprehensive insights into local coordination and structural changes.

## **Li Content Inference**

We directly applied the model trained on simulated data to real experimental data. The training dataset comprised 701 structures, with the spectrum of each element calculated from the site-averaged ensemble within the structure. The task of predicting Li content served as the validation basis, given that it is a quantifiable variable in experiments as the voltage profile reflects changes in Li concentration during cycling. Although direct measurement of Li content in experiments is challenging, the bulk averaged Li content can be inferred from the measured electrochemical capacity $C_m$ using:

$$Li\% = 1 - \frac{C_m}{C_t} \text{ and } C_t = \frac{nF}{MW}$$

where $C_t$ is the theoretical electrochemical capacity, n is the number of electrons per molecular weight (which is 1 here), F is the Faraday constant, and MW is the molecular weight.

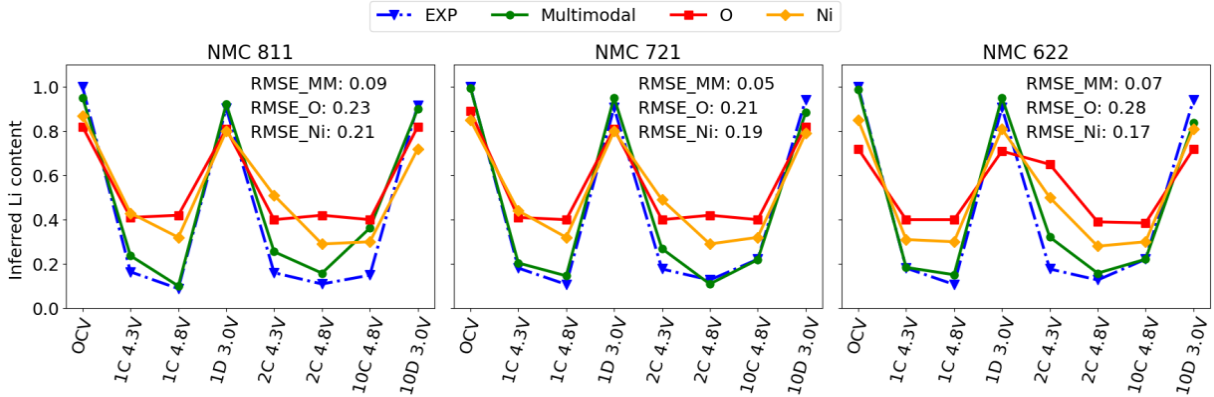

**Figure 3** Inference accuracy of Li content using multimodal (green) and unimodal approaches (O K-edge: red squares; Ni L-edge: red stars) compared with ground truth (blue) on experimental data of NMC811, NMC721 and NMC622. The X-axis represents the cycle and voltage states, starting from the open circuit voltage (OCV) and progressing through charge (C) and discharge (D) stages at specified voltages for the 1st, 2nd, and 10th cycles.

We adjusted the simulated data to align with the energy range of the experimental data by trimming and interpolating it accordingly. **Figure 3** illustrates the comparison between the predicted Li content and the "ground truth" derived from capacity measurements. Impressively, our model demonstrated exceptional prediction accuracy, with a root mean square error (RSME) value of less than 0.1. Such high accuracy is paramount in battery research, where precise estimation of Li content and capacity is crucial for understanding electrochemical processes and optimizing battery performance. Although bulk average Li content is known from electrochemical cycling data as indicated above, our approach allows the determination of local Li content from XANES or EELS mapping.

Moreover, our analysis highlighted the notable superiority of the multimodal approach over the unimodal approach in inferring Li content. By leveraging information from both O K-edge and Ni/Mn/Co $L_{2,3}$-edges, the multimodal approach achieved significantly enhanced predictive performance, reducing RMSE by 30-50% compared to single-edge inference. This finding underscores the importance of incorporating diverse spectroscopic features for comprehensive characterization. Even when considering Ni $L_{2,3}$-edges, renowned for their sensitivity to the redox reactions crucial in NMC, [37] the unimodal approach falls short compared to the multimodal strategy.

## Local Environment Inference

Understanding the local environment in NMC is critical for revealing the correlation between atomic-scale composition and electrochemical performance. Accurately identifying the local coordination of transition metals, lithium, and oxygen atoms allows for a more detailed analysis of how variations in local structure influence redox behavior, capacity retention, and material stability. By mapping the local environment around individual atoms, insights into how different elemental distributions impact the performance of NMC materials at the nanoscale can be achieved, which is essential for optimizing battery performance and longevity.

Considering the spatial resolution of EELS measurements in modern STEM techniques, typically ranging between 0.1 and 1 nm,[38] we defined the "local environment" as a spherical region with a radius of 0.3 nm. Within this region, the coordination shell typically contains three transition metal atoms, three oxygen atoms, and up to three lithium atoms. This local coordination configuration results in 24 possible scenarios, including combinations of up to three Li atoms with six different TM atom arrangements, leading to approximately 2800 possible local configurations. For each shell, the spectra of each element are calculated using site-average ensemble. In instances where an element (specifically for Mn and Co in our case) is absent within this shell, its spectrum is approximated by a constant line at y=0.1 with 2% Poisson noise added.

For this part, a single model was employed to predict four properties: the counts of Li, Ni, Mn, and Co atoms within the local coordination configuration. The comparative accuracies of multimodal and unimodal approaches are shown in **Figure 4(a)**. The multimodal strategy demonstrates substantial improvements in prediction accuracy, with margins ranging from 0.14 to 0.78 compared to its unimodal counterpart.

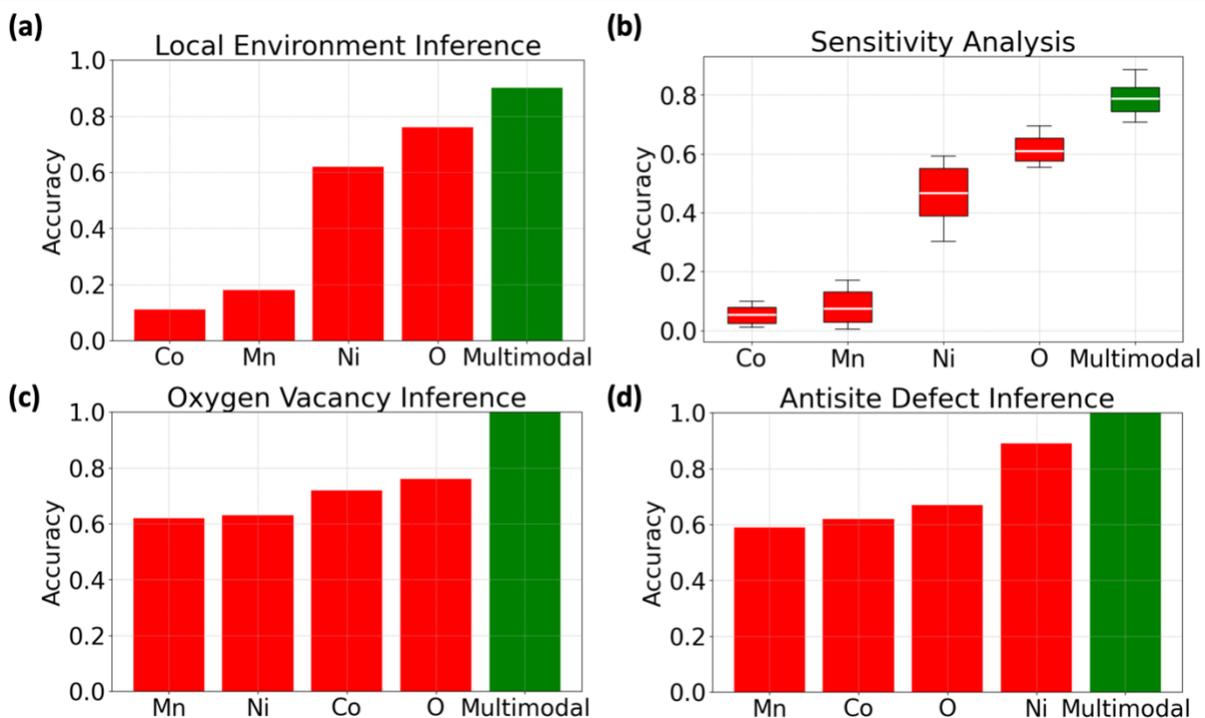

**Figure 4.** (a) Local environment inference accuracy using multimodal and unimodal approaches. (b) A Permutation Importance plot for multimodal and unimodal approaches applying 2% Poisson noise. The box represents the interquartile range of prediction accuracy, with edges marking the 25th and 75th percentiles and the internal line indicating the median value. The line error bar reflects the full range of prediction accuracies we observed. Local defect prediction accuracy using multimodal and unimodal approaches for (c) oxygen vacancies and (d) anti-site defects.

To further validate the robustness of these findings, a sensitivity analysis was conducted, focusing on the impact of noise interference. Given that Poisson noise is a predominant source in EELS/XAS data,[39,40] a 2% Poisson noise was applied to the spectral data of each element to mimic real-world conditions. This process was repeated over ten iterations to generate a comprehensive set of data reflecting various noise conditions. The results of this analysis are presented in a permutation importance plot, shown in **Figure 4(b)**, which highlights the comparative resilience of each element to noise interference. The findings clearly indicate that the multimodal approach is more robust against noise, outperforming the unimodal strategies in maintaining accuracy under these perturbed conditions.

## **Local defect**

Defects, such as oxygen vacancies and antisite defects, in NMC cathodes also play a critical role in determining the material's electrochemical performance and stability. Oxygen vacancies can influence charge transport and alter oxidation states, potentially improving capacity but also leading to structural degradation. Antisite defects disrupt the ideal atomic arrangement, reducing ionic conductivity and causing uneven charge distribution. Studying these defects is essential for optimizing NMC materials, as they can impact both the capacity retention and the long-term stability of the battery, ultimately guiding the development of more durable and efficient cathodes.

For oxygen vacancies, we used 136 defective structures and 701 pristine structures. Similarly, the spectrum of each element calculated from the site-averaged ensemble within the structure. To address the issue of class imbalance, low weights were assigned to pristine structures with higher Li content, as they lack corresponding defective structures.

Despite experimental evidence suggesting the sequence of local instability as Ni > Co > Mn in NMC structures,[41] our findings reveal that Co is more sensitive to oxygen vacancies than Ni, consistent with recent experimental observations for Ni-rich NMC materials.[42] However, since NMC materials are often designed to be Ni-rich to optimize the specific capacity,[43] the low concentration of Co in these materials poses challenges in identifying oxygen vacancies through Co alone. Nevertheless, employing a multimodal approach significantly enhances prediction accuracy, achieving 100% accuracy, as demonstrated in **Figure 4(c)**.

To evaluate our model, we directly applied the trained model to a new system with a different composition: 5% Si-doped Li-rich NMC. The annular dark field (ADF) image of the sample is shown in **Figure 5(a)**. The EELS data were acquired from the [100] surface. Given the distinct energy ranges for Si L-edges and O K-edge and TM L-edges, two separate acquisitions were conducted. In **Figure 5(b)**, pixels with Si L edges are labeled in green, and the predicted oxygen vacancies from O K-edge and TM L-edges are indicated in red. The significant overlap between these pixels suggested that Si promotes the formation of oxygen vacancies from TM–O bonds in proximity to the dopants, which is consistent with a previous study by Nation et al..[44] This overlap validated our model, even when applied to a new system incorporating not only additional Li in the TM layers, but also dopants.

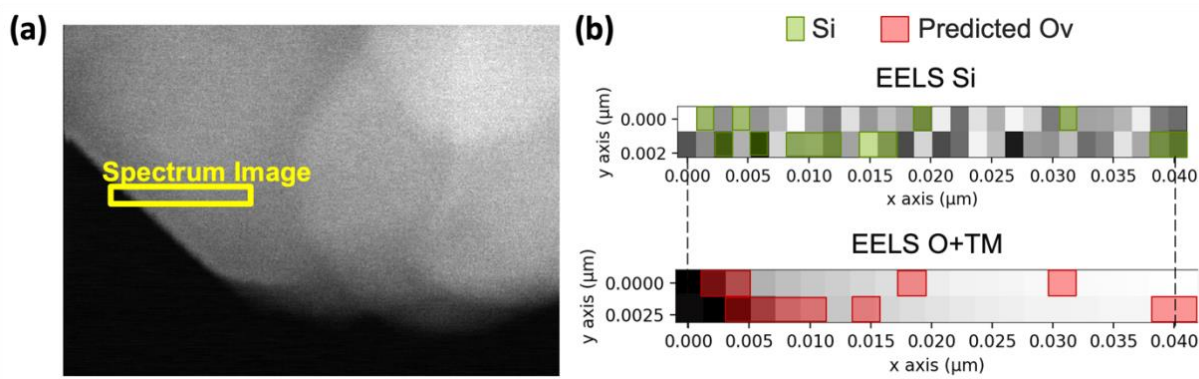

**Figure 5.** (a) ADF image of the 5% Si-doped Li-rich NMC. The EELS spectrum image was acquired from the window highlighted in yellow. (b) EELS maps acquired from (a). Pixels with Si L-edges are highlighted in green, while pixels with predicted oxygen vacancies are labeled in red.

Regarding anti-site defects, we evaluated 18 defective structures. The size of the dataset was constrained by the high computational cost associated with each calculation, necessitating the use of a 120-atom supercell. Additionally, many of the calculated configurations were found to be unstable, further limiting the available data. Our simulations show that the O K-edge offers limited insight when spatial resolution exceeds 0.3nm, largely due to the abundance of oxygen atoms within this spatial extent. Moreover, the sensitivity of Mn and Co to these defects is highly dependent on their proximity to the defect. Given the limited data on antisite defects and the sub-optimal results obtained at large spatial resolution, our analysis was refined to a 0.3 nm resolution. The findings, as shown in **Figure 4(d)**, reveal nickel's particular sensitivity to Ni/Li anti-site defects. However, similar to our approach with oxygen vacancies, leveraging a multimodal strategy significantly increases prediction accuracy, once again achieving a remarkable 100% accuracy rate.

## **Model Interpretation**

In our endeavor to ensure the transparency and interpretability of our models, we undertook a thorough analysis of feature importance. By employing ensembles of decision tree methods, such as XGBoost, we were able to assess the significance of each feature based on the number of splits within a trained predictive model. This enabled us to map the feature importance back to the original energy domain, as demonstrated in **Figure 6** for the prediction of Li content. Our results suggested that the Ni $L_3$ edges, which correspond to the excitation of $2p_{3/2}$ electrons to 3d orbitals, is the most important feature. This aligns with previous work, as such an excitation is particularly sensitive to Ni's oxidation states, thereby offering a direct correlation with the redox reaction and, consequently, to the Li content.[43] Similarly, the O K-edge was identified as the second most important feature, with both peaks playing vital roles: the first peak (corresponding to $\pi^*$) reflects the O-TM bonding while the second peak (corresponding to $\sigma^*$) relates more to charge-compensation mechanisms.[45,46] Through these findings, we demonstrated that our model has scientific interpretation, and our multimodal approach provides a direct comparison of the relative importance of each element in predicting various properties.

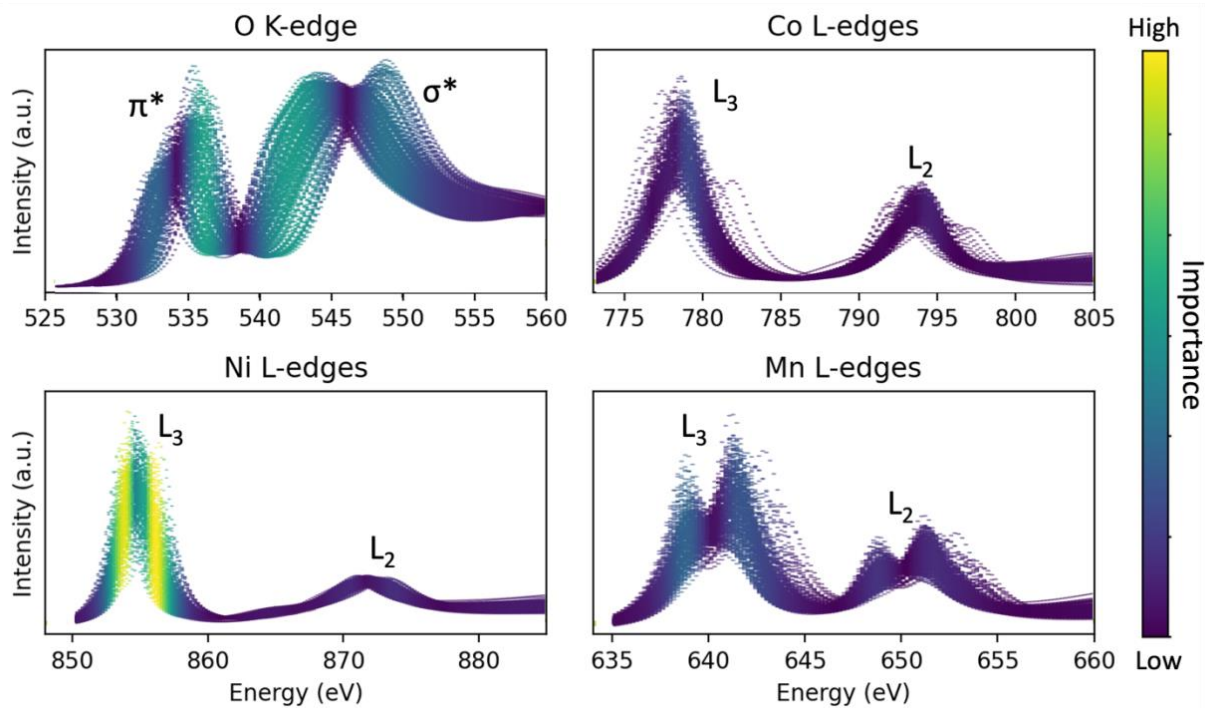

**Figure 6.** Feature importance plot for predicting Li content. Brighter colors indicate higher importance, while darker colors indicate lower importance.

# Conclusion and Future Work

In conclusion, our study highlighted the pivotal role of multimodal spectroscopy in advancing the characterization and understanding of complex materials, with NMC serving as a representative sample system. Through the integration of ab initio calculations, machine learning algorithms, and experimental data acquisition using multimodal data, we achieved profound characterization insights into local coordination and defect. Our results showed that multimodal approach not only enhanced the characterization accuracy, but also enabled robust cross-validation and facilitated the measurement of complex structures that may be challenging or even impossible to accurately capture using a single data source alone. Our sensitivity analysis further underscored the superior resilience of the multimodal approach to noise. Our results signified the transformative potential of multimodal spectroscopy in advancing the development of next-generation energy storage technologies. Moreover, this framework can be broadly applied to other material systems, accelerating materials discovery and innovation across various fields of research and development.

Looking ahead, there are several promising avenues for future research. Firstly, we can extend our model to analyze EELS hyperspectral time-series data. This approach will enable us to generate dynamic maps illustrating how specific properties evolve during reactions, such as the formation of defects and the diffusion of certain elements. Such dynamic insights are crucial for understanding the underlying mechanisms governing material behavior and can provide valuable guidance for optimizing materials performance. Next, the trained modal can be applied to new systems with some common elements, such as Na-NMC. By combining a limited amount of data from the new system and transfer learning, the model can be fine-tuned to accurately infer properties and behaviors in the new system. This approach will leverage the knowledge gained from the original NMC system to enhance the predictive capabilities for similar materials, thereby reducing the need for extensive new datasets and accelerating the development of novel materials. Furthermore, future research efforts can focus on integrating insights from XAS/EELS with complementary spectroscopic techniques, such as XRD and Raman spectroscopy. These techniques offer unique perspectives on different aspects of material structure and behavior, including long-range order and changes in crystallography. By combining various characterization methods, we can gain a more comprehensive understanding of material properties and behaviors, leading to enhanced insights for material design and development.

# Methods

## DFT and Spectra Simulations

All first principles structural calculations were performed using the VASP code[47], employing the projector-augmented wave (PAW) method.[48] For both pristine and oxygen-deficient structures, exchange-correlation effects were considered using the strongly constrained and appropriately normed (SCAN) functional.[49] To correct for the strongly correlated $d$ electrons, the Hubbard $U$ correction was applied, with $U$ values set to 2.43 eV for Ni, 2.93 eV for Mn, and 2.86 eV for Co.[50] In simulations involving antisite defects, the exchange-correlation was treated using the Perdew–Burke–Ernzerhof (PBE) generalized gradient approximation (GGA), based on literature specific to NMC materials with antisite defects.[51] Correspondingly, Hubbard $U$ values were adjusted to 6.7 eV for Ni, 4.2 eV for Mn, and 4.9 eV for Co.[52] Initial magnetic configurations were set to be ferromagnetic, with Co in a low spin state and both Ni and Mn in high spin states, aligning with findings from prior research.[53,54] The plane wave energy cutoff was set to 450 eV, and convergence thresholds were established at $1\times10^{-4}$ eV for energy and 0.05 eV/Å for force.

All the systems studied in this work are layered structures with a space group $R\bar{3}m$, modeled after LiCoO$_2$ (structure prototype α-NaFeO$_2$). To construct the NMC compositions, Co is replaced with an appropriate mixed occupancy of Ni:Mn ratios to achieve NMC622, NMC811, and NMC721 compositions. Additionally, the Li site is partially occupied to represent lithiation levels in intervals of 0.1. Symmetrically distinct orderings of transition metals and Li/vacancy configurations are enumerated using Pymatgen.[55] For pristine structure calculations, supercells containing 30-40 atoms, comprising 0-10 Li atoms, 10 transition metal (TM) atoms, and 20 oxygen atoms were used. For oxygen defect studies, supercells containing 29-31 atoms were employed, incorporating a 5% oxygen vacancy rate and 0-20% Li content, reflecting the increased likelihood of oxygen vacancies at lower Li concentrations, as supported by Zheng et al. (2016), Li et al. (2020), and Jung et al. (2017).[56-58] For antisite defect calculations, larger 120-atom supercells were considered, consisting of 30 Li atoms, 30 TM atoms, and 60 O atoms, with 3% Ni/Li anti-sites, as discussed by Zheng *et al.* (2017).[59]

Regarding spectroscopy, although XAS and EELS both yield very similar experimental spectra and are based on the excitation of inner shells, their underlying principles differ. The key difference between these two methods lies in the double-differential cross-section.[60] However, when the absorption energy exceeds 100 eV, such as those for the O K-edge and Ni/Mn/Co $L_{2,3}$-edges considered in this work, this difference becomes negligible.[61] Therefore, simulation codes for both EELS and XAS were considered in benchmark tests. Specifically, the full-potential finite difference method from FDMNES[62], plane-wave pseudopotential method from XSPECTRA (Quantum ESPRESSO)[63,64], and VASP[47], the Bethe-Salpeter equation method from OCEAN[65,66], a multiple scattering approach from FEFF[67], and an all-electron full-potential linearized augmented-plane-wave method from ELK[68] were considered.

To quantitatively evaluate the simulation results, cosine similarity and Pearson correlation metrics were used for comparison against experimental data. The benchmarking results using NMC333 are illustrated in **Figure 7**, which showcase the comparative performance of each simulation code against the experimental benchmarks. Among the computational methods outlined in the **DFT and Spectra Simulations** section, FDMNES and FEFF8 demonstrated the highest agreement with the experimental data for the O K-edge. Regarding the transition metal $L_{2,3}$-edges, FEFF exhibited similar evaluation scores with FDMNES, with slightly higher cosine similarity but slightly lower Pearson correlation. However, FDMNES more accurately reproduced the $L_3/L_2$ peak ratios and retained greater detail in the spectral features, such as the definition of peak shoulders. Therefore, FDMNES was selected for simulating all spectra in this study.

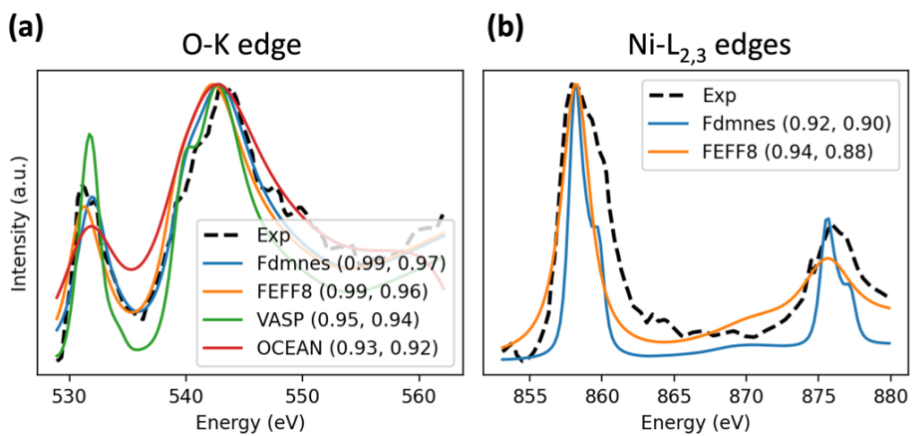

**Figure 7.** Benchmarking results for the O K-edge (a) and Ni L$_{2,3}$-edges (b) in pristine NMC333, comparing experimental data [73] (black dotted line) with simulations from FDMNES (blue), FEFF8 (orange), VASP (green), and OCEAN (red). For each code, the cosine similarity and Pearson correlation scores are labeled after their names.

## Machine Learning Model Evaluations

The prediction accuracy is defined as

$$accuracy = \frac{TP_{total} + TN_{total}}{TP_{total} + TN_{total} + FP_{total} + FN_{total}}$$

where True Positive (TP) represents the correctly predicted count of a specific atom type (e.g., Li, Ni, Mn, Co) within a given region. False Positive (FP) occurs when the predicted count of an atom type exceeds the actual count, and False Negative (FN) occurs when the predicted count of an atom type is less than the actual count. True Negative (TN) is 1 when the absence of an atom type is correctly predicted and 0 otherwise. The term "total" refers to the summation across all elements considered.

The benchmarking results using different models are shown in **Table 1**:

| Model | Li | TM |
|---|---|---|
| **RIDGE** | 0.93 | 0.79 |
| **MLP** | 0.99 | 0.84 |
| **RF** | 0.95 | 0.88 |
| **XGBOOST** | 0.99 | 0.93 |
| **GBM** | 0.96 | 0.84 |
| **LGBM** | 0.99 | 0.88 |
| **CNN** | 0.97 | 0.77 |

**Table 1**. Comparison of various machine learning models used for local environment inference. The table reports the accuracy of each model for inferring Li content and TM content, with models including Ridge regression, Multi-Layer Perceptron (MLP), Random Forest (RF), XGBoost, Gradient Boosting Machine (GBM), LightGBM (LGBM), and Convolutional Neural Networks (CNN).

## Featurization and Dimensionality Reduction

In this study, we employed four methods to featurize the spectra shown in **Figure 8**. Firstly, the raw intensity feature comprises paired energy and intensity values. Secondly, we utilized the Cumulative Distribution Function (CDF) of the raw intensity[24]. Thirdly, we implemented peak descriptors based on the methodology proposed by Torrisi *et al.*[69], where the energy range is divided into N equal regions. Within each region, a cubic polynomial is fitted to the spectrum, with region sizes set at 2.5 eV, 5 eV, and 10 eV, resulting in four coefficients per region. Lastly, we decomposed the spectra into 20 Gaussian peaks, each characterized by three parameters: center, amplitude, and width (full width at half maximum limited between 0 and 5 eV).

For dimensionality reduction, we explored three techniques for each feature set: Truncated Singular Value Decomposition (SVD) and Principal Component Analysis (PCA) as linear decomposition methods, and Isomap as a non-linear technique. The optimal number of dimensions for each reduced dataset was determined by selecting the minimum number of dimensions required to explain at least 90% of the total variance.

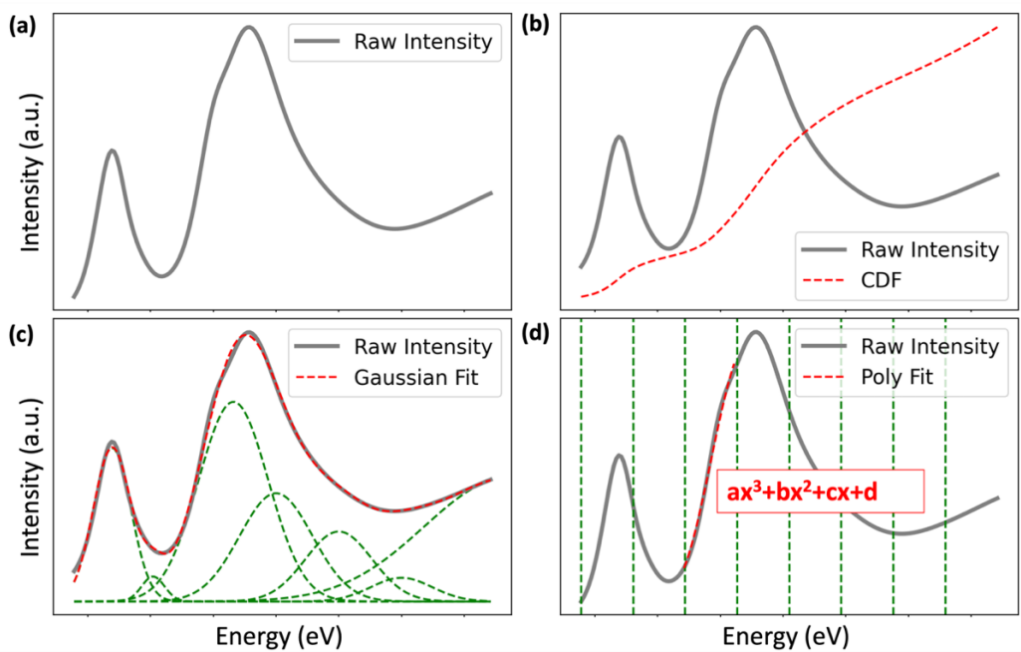

**Figure 8**. Visualization of spectral featurization for (a) raw intensity, (b) CDF, (c) Gaussian fit, and (d) polynomial fit.

## Model Construction and Training

The selection of ML algorithms encompassed various techniques, including Ridge regression, multi-layer perceptron (MLP), random forest (RF), extreme gradient boosting machine (XGBoost), gradient boosting machine (GBM), light gradient boosting machine (LightGBM), and convolutional neural networks (CNN).

Training ML models is a time-intensive process that significantly impacts outcomes. To enhance efficiency and efficacy, we employed Bayesian Optimization for hyperparameter tuning. Unlike grid or random searches, Bayesian Optimization utilizes historical results to inform decisions on the next set of hyperparameters to evaluate. This approach typically requires fewer iterations to identify optimal or near-optimal hyperparameters, which is particularly beneficial in scenarios with limited data or computational resources. By minimizing the number of evaluations, Bayesian Optimization optimizes the model-tuning process, make it particular useful for cases with limited data points.

While Gaussian processes are commonly used as the surrogate model for Bayesian Optimization, scalability issues arise due to their cubic computational complexity with the number of data points.[36] To address this, we utilized the Tree-structured Parzen Estimator (TPE). The Probability of Improvement (PI) is defined as:

$$PI(x) = P(f(x) \geq \gamma) = \int_{\gamma}^{\infty} p(f(x)) df(x)$$

where $f(x)$ represents the objective function value at point $x$, and $\gamma$ is the best observed value so far. The Expected Improvement (EI) is given by:

$$EI(x) = \mathbb{E}[(f(x) - \gamma)^+] = \int_{0}^{\infty} [(f(x) - \gamma)^+] p(f(x)) df(x)$$

where $(f(x) - \gamma)^+$ denotes the positive part of $f(x) - \gamma$. BO-TPE constructs two separate probability density functions (PDFs) for "good" and "bad" hyperparameters, denoted as $p(\theta|y < \gamma)$ and $p(\theta|y > \gamma)$ respectively, where $\theta$ represents the hyperparameters and $y$ represents the evaluation metric. For our classification models, the evaluation metric was Softmax, whereas for regression models, it was the squared error. The ratio of these PDFs determines the acquisition function, guiding the suggestion of new hyperparameter configurations to efficiently explore and exploit the search space. TPE demonstrated superior convergence compared to other surrogate models.[36,70-72]

## Experimental measurements

### Soft X-ray absorption spectroscopy (sXAS)

O $K$-edge and Ni $L_3$-edge sXAS spectra of NMC622, NMC721, NMC811 are collected at beamline 8.0.1.1 in Advanced Light Source.[74] The electrodes are casted on Al foil with the composition of 92:1:7, active materials: carbon additive (acetylene black) : binder (KF1100, Kureha), respectively. 2032 type coin cells are assembled with 1 M LiPF$_6$ EC:DEC (1:1 v/v) electrolyte and charged to various states of charges. The charged electrodes are dissembled in the glovebox and transferred to measurement chamber by using air-tight transfer kit.

### EELS

Li$_{1.2}$Mn$_{0.49}$Si$_{0.05}$Ni$_{0.13}$Co$_{0.13}$O$_2$ powder was synthesized using the co-precipitation method described previously.[75] A slurry of 80 : 10 : 10 active material: PVDF binder: conductive carbon was coated at a 10 mil wet thickness onto aluminum foil and dried at 80°C. Coin cells were assembled in an argon filled glovebox with lithium metal counter electrode, Celgard 2325 separator and 1 M LiPF$_6$ in 1/1 EC/DEC electrolyte. First cycle voltage measurements were conducted at C/100 and rate capability cycling conditions ranged from C/20–2C, where the 1C current corresponds to approximately 2 mA and 1.6 mg cm$^{-2}$ loading.

For transmission electron microscopy (TEM) characterization, cells were charged at a C/10 rate to 4.4 V (cutoff voltage before activation) and 4.6 V (cutoff voltage after activation) and disassembled after the first charge cycle. Charged electrodes were harvested in the glovebox, rinsed in DMC, and dried. The dried electrode was then scraped off the current collector for STEM and EELS post-mortem characterization. Sample powders were loaded onto a lacy carbon grid for TEM analysis. STEM imaging and EELS were performed with an aberration corrected FEI Titan microscope at 300 kV equipped with a Gatan Image Filter Quantum-865. HAADF-STEM images were recorded with a probe convergence angle of 30 mrad and a large inner collection angle of 65 mrad. The EELS background was removed using Digital Micrograph software through a power-law fitting in the pre-edge region of each spectrum using a consistent background window.

## Acknowledgements

This work is funded by the Energy Storage Research Alliance "ESRA" (DE-AC02-06CH11357), an Energy Innovation Hub funded by the U.S. Department of Energy, Office of Science, Basic Energy Sciences. M.K.Y.C. acknowledges the support from the BES SUFD Early Career award. Work performed at the Center for Nanoscale Materials, a U.S. Department of Energy Office of Science User Facility, was supported by the U.S. DOE, Office of Basic Energy Sciences, under Contract No. DE-AC02-06CH11357. This research used resources of the National Energy Research Scientific Computing Center, a DOE Office of Science User Facility supported by the Office of Science of the U.S. Department of Energy under Contract No. DE-AC02-05CH11231. We gratefully acknowledge the computing resources provided on Bebop, a high-performance computing cluster operated by the Laboratory Computing Resource Center at Argonne National Laboratory. Soft X-ray experiments were performed at BL8.0.1 of the Advanced Light Source (ALS), a DOE Office of Science User Facility, under contract no. DE-AC02-05CH11231.